\documentclass[superscriptaddress]{revtex4}
\usepackage{amssymb}  
\usepackage{amsfonts} 
\usepackage{amstext}  
\usepackage{graphicx}							
\usepackage[dvips]{color} 
\usepackage{epsfig,amsmath,lscape}
\begin{document}

\title{The importance of strange mesons in neutron star properties}

\author{R. Cavagnoli}
\affiliation{Depto de F\'{\i}sica - CFM - Universidade Federal de Santa
Catarina - Florian\'opolis - SC - CP. 476 - CEP 88.040 - 900 - Brazil}
\author{D.P.Menezes}
\affiliation{Depto de F\'{\i}sica - CFM - Universidade Federal de Santa
Catarina - Florian\'opolis - SC - CP. 476 - CEP 88.040 - 900 - Brazil}

\begin{abstract}
In order to obtain the properties of compact stellar objects, appropriate equations of state have to be used. In the literature, strange meson fields, namely the scalar meson field ${\sigma}^*(975)$ and the vector meson field $\phi(1020)$, had to be considered in order to reproduce the observed strongly attractive $\Lambda\Lambda$ interaction. The introduction of these strange mesons makes the equations of state harder (EOS) due to the repulsive effect of the $\phi(1020)$ meson. In this work the inclusion of these mesons in the equation of state and their influence on the properties of the neutron stars are investigated.
\end{abstract}

\maketitle

\section{Introduction}

In the present work we use the relativistic non-linear Walecka model (NLWM) \cite{walecka}, at zero temperature ($T = 0$), with the lowest baryon octet $\left\{ N, \Lambda, \Sigma , \Xi  \right\}$ in $\beta$ equilibrium with the lightest leptons $\left\{ e^- , {\mu}^- \right\}$ and compare the results with the same model plus strange meson fields, $\sigma ^* (975) 
$  and  $\phi(1020)$, which introduce strangeness to the interaction according to \cite{alemao} and \cite{greiner}. Strange meson fields, namely the scalar meson field $\sigma ^* (975)$ and the vector meson field $\phi(1020)$, had to be considered in order to reproduce the observed strongly attractive $\Lambda\Lambda$ interaction. This formalism applied to compact objects like neutron stars, where the energies are such that allow the appearance of the eight lightest baryons. The motivation for this study lies in our interest to describe the interaction between hadrons taking into account a growing number of effects in order to better describe it. Given the difficulty of making comparisons with experimental data we will, for now limit ourselves to verify if the inclusion of these mesons significantly alters some
quantities like pressure and energy density in the equation of state and the bulk properties of compact stars.

\section{The Formalism}

The lagrangian density of the NLWM with the inclusion of the strange meson 
sector and leptons for $\beta$ equilibrium is:

$$ {\cal L} = \sum\limits_{B{\kern 1pt}  = {\kern 1pt} 1}^8 {\bar \Psi _B \left[ {\gamma _\mu  \left( {i\,\partial ^\mu - g_{\omega{\kern 1pt} B} \,V^\mu - g_{\rho {\kern 1pt} B} \vec \tau  .\vec b^\mu  } \right) - \left( {M_B - g_{\sigma{\kern 1pt} B} \sigma } \right)} \right]\Psi _B }$$
$$+ \frac{1}{2} \left( {\partial _\mu \sigma \partial ^\mu \sigma - m_{\sigma}^2 \sigma ^2 } \right) - \frac{1}{{3!}} k \sigma ^3  - \frac{1}{{4!}} \lambda \sigma ^4 - \frac{1}{4} \Omega _{\mu \nu } \Omega ^{\mu \nu } + \frac{1}{2} m_{\omega}^2 V_\mu  V^\mu $$
$$- \frac{1}{2} \vec B_{\mu \nu } \vec B^{\mu \nu } + \frac{1}{2} m_\rho ^2 \vec b_\mu  \vec b^\mu $$
$$+\frac{1}{2} \left( {\partial _\mu \sigma ^*  \partial ^\mu  \sigma ^*  - m_{\sigma^*}^2 \sigma ^{*2} }\right) + \frac{1}{2} m_\phi^2 \phi _\mu  \phi ^\mu  - \frac{1}{4} S_{\mu \nu } S^{\mu \nu }
- \sum\limits_B {{\kern 1pt} g_{\sigma ^*  {\kern 1pt} B} \bar \Psi _B \Psi _B {\kern 1pt} \sigma ^*} - \sum\limits_B {g_{\phi {\kern 1pt} B} \bar \Psi _B \gamma _\mu {\kern 1pt} \Psi _B {\kern 1pt} \phi ^\mu} $$
\begin{equation}
+\sum\limits_{l{\kern 1pt}  = {\kern 1pt} 1}^2 {\bar \Psi _l \left( {i\gamma _\mu  \partial ^\mu - M_l } \right)\Psi _l }, \\ 
\end{equation}
\noindent where
$\Omega _{\mu \nu } =\partial _\mu V_\nu - \partial _\nu V_\mu $, 
$\vec B_{\mu \nu } = \partial _\mu  \vec b_\nu - \partial _\nu \vec b_\mu - g_\rho  \left( {\vec b_\mu \,\times \, \vec b_\nu } \right)$ and
$S_{\mu \nu } = \partial _\mu \phi _\nu  - \partial _\nu \phi _\mu $,
with $B$ extending over the eight baryons, $g_{iB}$ are the coupling constants of mesons $i$,~$i = \sigma, \omega, \rho$ with baryon $B$, and $m_i$ is the mass of meson $i$. $\lambda$ and $k$ are the weighs of the non-linear scalar terms and $\vec \tau$ is the isospin operator. 
At this point it is worth emphasizing that the strange mesons are not supposed
to act at low densities, where the strangeness content is zero. Moreover, the 
non-linear terms are normally corrections added to the main linear 
contributions and hence the non-linear terms in the strange sector are 
disregarded in the present work.
The constants $g_{iB}$ are defined by $g_{iB} = x_B\,g_i$ where $x_B = \sqrt{2/3}$, for hyperons, \cite{mos}, $x_B = 1$ for the nucleons, and also $g_{\sigma} = 8.910$, $g_{\omega} = 10.626$, $g_{\rho} = 8.208$, $g_{\sigma ^* \Lambda } = g_{\sigma ^* \Sigma} = 5.11$, $g_{\sigma ^* \Xi} = 9.38$, $g_{\phi \Lambda} = g_{\phi \Sigma} = 4.31, g_{\phi \Xi} = 8.62$, $k = - {\rm 6}{\rm .426} \;10^{-4}\,,\; \lambda = 5.530$ according to \cite{gle1} and \cite{gle2}. 
The strange mesons interact with hyperons only ($g_{\sigma ^* {\kern 1pt} p} = g_{\sigma ^* {\kern 1pt} n} = g_{\phi {\kern 1pt} {\kern 1pt} p} = g_{\phi {\kern 1pt} {\kern 1pt} n} = 0$).
The masses of baryons of the octect are: $M_N = 938\,MeV$ (nucleons), $M_{\Lambda} = 1116\,MeV$, $M_{\Sigma} = 1193\,MeV$, $M_{\Xi} = 1318\,MeV$ and the meson masses are: $m_\sigma = 512\,MeV$, $m_\omega = 738\,MeV$, $m_{\rho} = 770\,MeV$, $m_{\sigma^*} = 975 \,MeV$, $m_\phi = 1020\,MeV$. In order to account for the $\beta$ equilibrium in the star the leptons are also included in the lagrangian density of eq. (1) as a non-interacting Fermi gas. The masses of the leptons are $M_{e^-} = 0.511\,MeV$ and $M_{\mu^-} = 105.66\,MeV$.

Applying the Euler-Lagrange equations to (1) and using the mean-field approximation ($\sigma \to \langle \sigma \rangle = \sigma _{{\kern 1pt} 0} ,\;V_\mu \to \langle V_\mu  \rangle = \delta _{\mu {\kern 1pt} {\kern 1pt} 0} \,V_0$~and~$\vec b_\mu  \to \langle \vec b_\mu  \rangle = \delta _{\mu {\kern 1pt} 0} \,b_\mu^0  \equiv \delta _{\mu {\kern 1pt}0} \, b_\mu $), we obtain:

\begin{equation}
	\sigma _{{\kern 1pt} 0} = - \frac{k}{{2 m_{\sigma}^2 }} \sigma _{{\kern 1pt} 0}^2 - \frac{\lambda }{{6 m_{\sigma}^2 }} \sigma _{{\kern 1pt} 0}^3 + \sum\limits_B^{} {\frac{{g_{\sigma}}}{{m_{\sigma}^2 }} x_{B} \rho _{{\sigma}{\kern 1pt} B} \; ,} 
\end{equation}
  
\begin{eqnarray}
&&V_{{\kern 1pt} 0} = \sum\limits_B^{} {\frac{{g_\omega }}{{m_\omega^2 }} x_{B} \rho _B \; ,} 
\\
\nonumber \\
&&b_{{\kern 1pt} 0} = \sum\limits_B^{} {\frac{{g_\rho  }}{{m_\rho ^2 }} x_{B} \tau _3 \rho _B \; ,} 
\end{eqnarray}  

\begin{eqnarray}
&& \sigma_{0}^* = \sum\limits_B^{} {\frac{g_{\sigma^*}}{m_{\sigma^*}^2}} 
x_B \rho_B \; ,
\\
\nonumber \\
&& \phi_{0} = \sum\limits_B^{} {\frac{g_\phi}{{m_\phi ^2 }} x_B \rho _B \; ,} 
\end{eqnarray}

\noindent where

\begin{equation}
\rho _{\sigma{\kern 1pt} B} = \frac{{M_B^ *  }}{{\pi ^2 }} \int_0^{K_{F_B}} {\frac{{p^2 dp}}{{\sqrt {p^2 + M_B^ *  } }}} \; ,
\end{equation}

\begin{equation}
 \rho _B = \frac{1}{{3 \pi ^2 }} K_{F_B}^3 \quad , \quad
M_B^* = M_B - g_{\sigma B} ~\sigma _{0} - g_{\sigma^* B}~\sigma _{0}^* \;,
\end{equation}  
and the 0 subscripts added to the fields mean that a mean field approximation, where the meson fields were considered as classical fields was performed.

\vspace{0.50cm}
Through the energy-momentum tensor, we obtain:

\begin{equation}
\varepsilon_a = \frac{1}{\pi^2} \left( {\sum\limits_{i = B,l}^{} {\int_0^{K_{F_i}} {p^2 dp \sqrt {p^2 + M_i^{*2}}}}} \right) + \frac{{m_\omega^2 }}{2} V_0^2 + \frac{{m_\rho ^2 }}{2} b_0^2 + \frac{{m_\sigma^2 }}{2} \sigma _{0}^2 + \frac{k}{6} \sigma _{0}^3 + \frac{\lambda }{{24}} \sigma _{0}^4
+ \frac{m_{\sigma^*}^2}{2} \sigma_{0}^{*2} + \frac{{m_\phi ^2}}{2} \phi_0^2,
\label{ea}
\end{equation}

\begin{equation}
P_a = \frac{1}{3 \pi^2} \left( {\sum\limits_{i = B, l}^{} {\int_0^{K_{F_i} }} {\frac{p^4 dp}{\sqrt {p^2 + M_i^{*2}}}}} \right) + \frac{{m_\omega^2 }}{2} V_0^2 + \frac{{m_\rho ^2 }}{2} b_0^2 - \frac{m_\sigma^2}{2} \sigma_{0}^2 - \frac{k}{6} \sigma _{0}^3 - \frac{\lambda }{24} \sigma _{0}^4 
- \frac{{m_{\sigma^*}^2 }}{2} \sigma _{{\kern 1pt} 0}^{ * {\kern 1pt} 2} + \frac{{m_\phi ^2 }}{2} \phi _{{\kern 1pt} 0}^{{\kern 1pt} 2} \;.
\label{Pa}
\end{equation}

In a neutron star, charge neutrality and baryon number must be conserved 
quantities. Moreover, the conditions of chemical equilibrium hold. In terms
of the chemical potentials of the constituent particles, these conditions 
read:

\begin{eqnarray}
&& \mu_{n} = \mu_{p} + \mu_{e^-} \; , \; \mu_{\mu^-} = \mu_{e^-}, \nonumber
\\
&& \mu_{\Sigma^0} = \mu_{\Xi^0} = \mu_{\Lambda} = \mu_{n}  \; , \nonumber
\\
&& \mu_{\Sigma^-} = \mu_{\Xi^-} = \mu_{n} + \mu_{e^-}  \; , \nonumber
\\
&& \mu_{\Sigma^+} = \mu_{p} = \mu_{n} - \mu_{e^-}  \; .
\end{eqnarray}

\section{Results}

\begin{figure}[ht]
  \centering
  \includegraphics[width=9cm,angle=0]{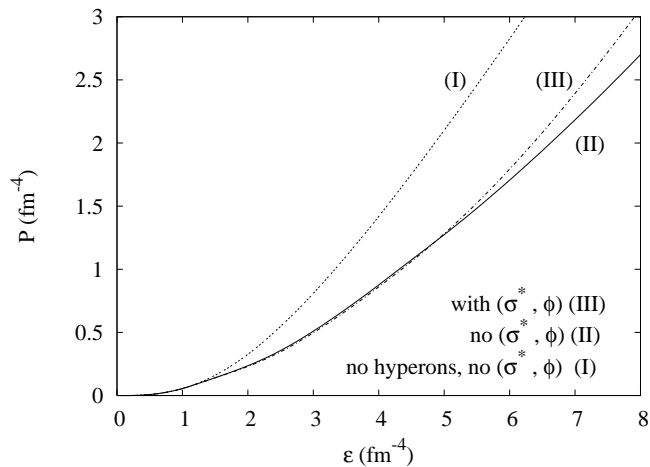}
  \caption{Pressure versus energy density.}
  \label{enerpress}
\end{figure}

\begin{figure}[ht]
  \centering
\begin{tabular}{cc}
\includegraphics[width=9cm,angle=0]{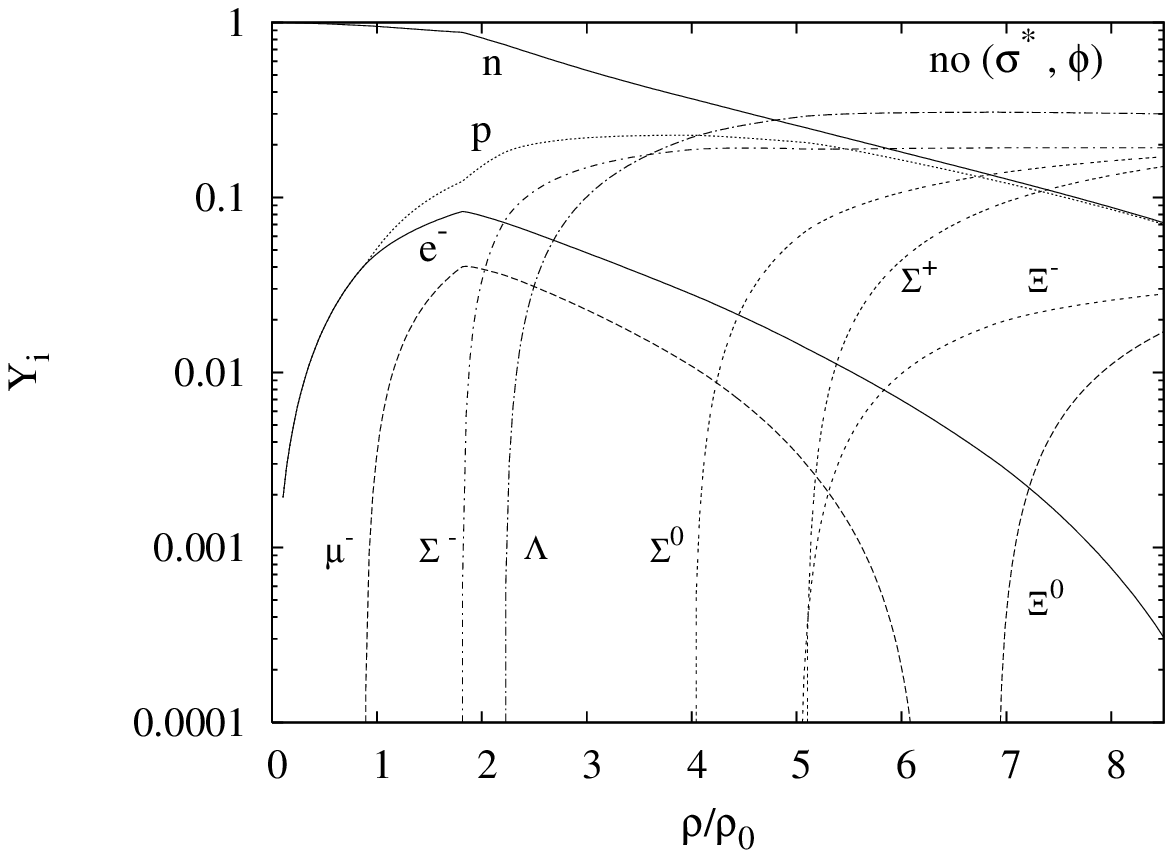}&
\includegraphics[width=9cm,angle=0]{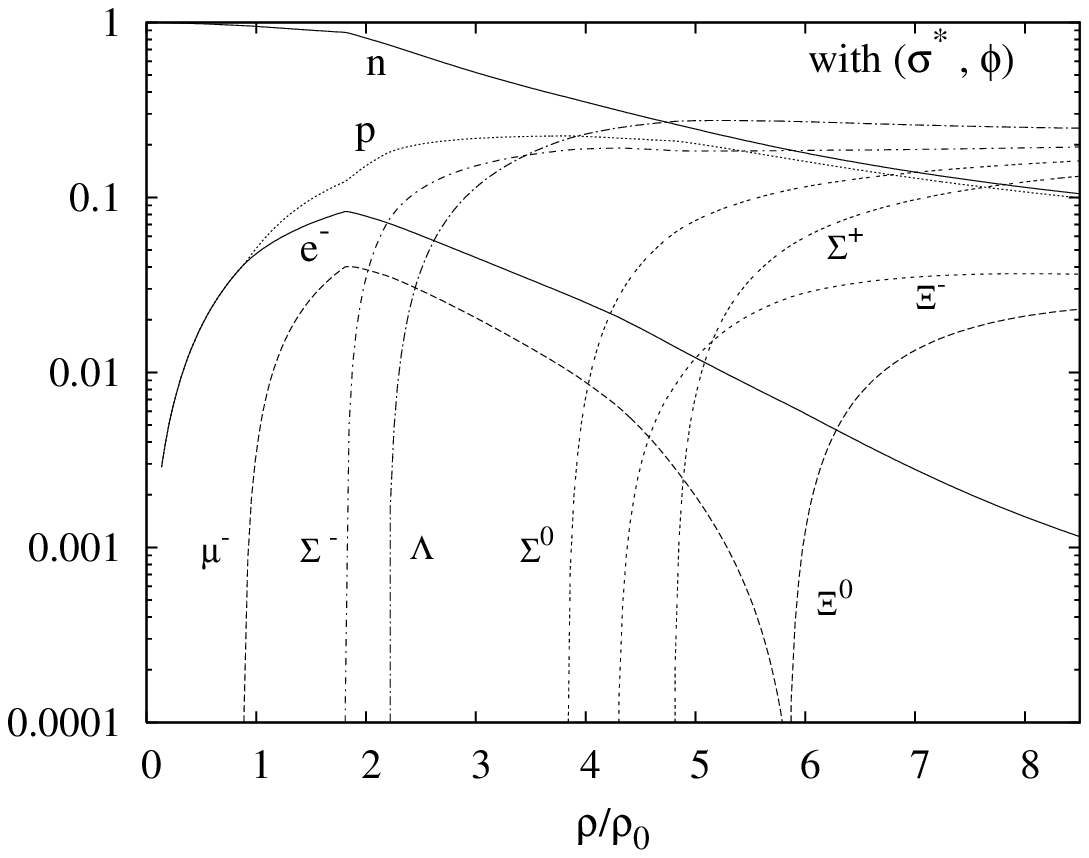}\\
\end{tabular}
 \caption{Particle fractions $Y_i = \rho_i /\rho$,\; i = B,l, as a function of the total baryon density $\rho$ in units of $\rho_0$.}
 \label{Yi}
\end{figure}

The inclusion of hyperons and strange mesons alters the equations of state
and the particle fractions, as can be seen from figures \ref{enerpress} and
\ref{Yi}.

From fig.~\ref{enerpress} we notice that the inclusion of the hyperons softens
the equations of state in comparison with the EOS obtained only with nucleons
and leptons. The inclusion of the strange mesons hardens these equations a 
little at higher energy densities. This indicates that the influence of the strange mesons is significant at higher densities, what can be easily seen in fig.~\ref{Yi}, where we notice a difference in the fractions of heavier hyperons, at densities above $5 \rho_0$, where $\rho_0$ is the saturation density of the nuclear matter.   

\begin{figure}[ht]
  \centering
  \includegraphics[width=9cm,angle=0]{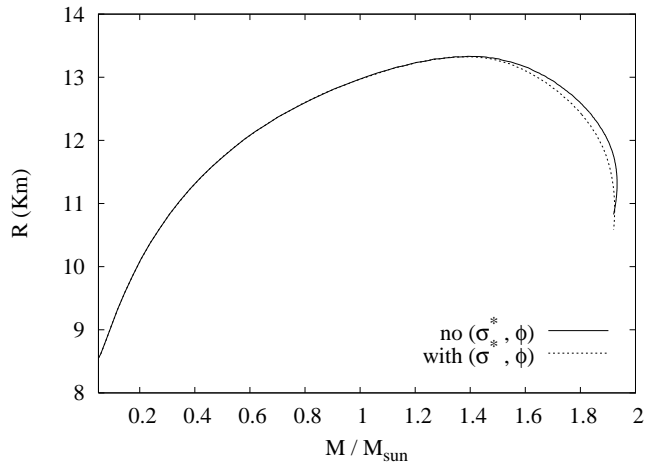}
 \caption{Radius versus gravitational mass for the families of stars obtained
within the NLWM and within the NLWM plus the strange mesons sector after
the TOV equations were solved.}
 \label{figtov}
\end{figure}

Neutron star profiles can be obtained by solving the Tolman-Oppenheimer-Volkoff (TOV) equations \cite{tov}, resulting from the exact solution of Einstein's general relativity equations in the Schwarzschild metric for spherically symmetric, static stars. Applying the equation of states (\ref{ea}) and (\ref{Pa}) 
in TOV equations results in the star properties shown in table I and figure
\ref{figtov}. In table I the profiles of the stars with the maximum 
gravitational mass and with the maximum radius are shown for two possible
EOS: without the strange mesons and with them. In these cases the crust of the 
stars were not included.

\newpage

\begin{table}[h]
\begin{tabular}{lcccccccc}
\hline
 & No ($\sigma^{*} , \phi$) & With ($\sigma^{*} , \phi$)\\
\hline
$M_{max}/M_{sun}$ & 1.93 & 1.92 \\
$R$ (Km) &  11.30 &  10.91 \\
$\varepsilon_0$ (fm$^{-4}$) & 6.39 & 7.01 \\
\hline
$R_{max}$ (Km) & 13.33 & 13.33 \\ 
$M/M_{sun}$ & 1.39 &  1.39 \\
$\varepsilon_0$ (fm$^{-4}$) & 1.75 & 1.78 \\
\hline
\end{tabular}
\caption{Neutron star properties. $\varepsilon_{0}$ is the central energy 
density.}
\end{table}

The observed values for the mass of the neutron stars lie between 1.2 to 1.8
$M_{sun}$. Our results are in the expected range. From table I and figure 3,
one can see that the differences in the star properties with and without strange meson are not very relevant. Nevertheless, the constitution of the stars at large densities are somewhat different. At about four times the saturation density
(see figure \ref{Yi}) the inclusion of the strange mesons start playing its 
role in the constitution of the stars. At this high energy a phase transition
to a deconfined phase of quarks or to a system with kaon condensates can 
already take place. These two possibilities are certainly more important
to the properties of neutron stars than a system containing strange mesons.
The influence of the inclusion of the strange mesons in protoneutron stars 
with temperatures around 30 to 40 $MeV$ and the their
importance when trapped neutrinos are included are under investigation.

\section*{ACKNOWLEDGMENTS}

This work was partially supported by CNPq (Brazil).

\end{document}